**Rolling of an elastomeric cylinder: a Marangoni like effect in solid**

Subrata Mondal[1] and Animangsu Ghatak[1,2,*]

[1] Department of Chemical Engineering, Indian Institute of Technology, Kanpur, 208016 (India)
[2] Center for Environmental Science and Engineering, Indian Institute of Technology, Kanpur, 208016 (India)
[*] Prof. A. Ghatak, Corresponding-Author, Author-Two, E-mail: aghatak@iitk.ac.in
Department of Chemical Engineering, Indian Institute of Technology, Kanpur, UP 208016

**Abstract:**

A soft, thin, elastomeric micro-cylinder is induced to roll on a solid substrate by releasing small quantity of a solvent. The solvent swells the cylinder asymmetrically at one side and evaporates out of it from where it is exposed to atmosphere. Because of inhomogeneous swelling, the cylinder bends but tends to straighten out following evaporation of the solvent. Balance of these two opposing effects induces the static state of the cylinder to eventually bifurcate to a dynamic state of rolling. Similar to marangoni effect in liquid, the rolling motion of cylinder is driven and sustained by the curvature of the cylinder. The rolling velocity increases linearly with curvature of the bent cylinder which ceases to locomote as the curvature diminishes below a threshold limit. Similar to marangoni effect, the rolling velocity increases also with temperature of the substrate and surroundings. A scaling relation derived for the rolling velocity captures all these observations.

**Introduction:**

Imbibition of a liquid into the crosslinked network of a gel or an elastomer is important for several scientific and technological applications, e.g. storage and delivery of drugs and nutrients (Hennink *et al.* 2002, Berger *et al.* 2004), soft actuation (Osada *et al.* 1992, Lu *et al.* 2000, Yuan *et al.* 2013), tissue engineering (Ng *et al.* 2005), design of micro-chemo-mechanical devices (Ueoka *et al.* 1997, Arora *et al.* 2009), variety of membrane separation processes e.g. water purification via reverse-osmosis (Idris *et al.* 2002, Chen *et al.* 2002) and even in mundane observations, like curling of a wet paper (Reyssat *et al.* 2011). The phenomenon is prevalent also in several biological settings, e.g. directed tissue growth leading to morphogenesis (Kao *et al.* 1992), actuation of hygromorphs (Edwards *et al.* 2005, Dawson *et al.* 1997), shape change in articular cartilage (Setton *et al.* 1998). In essence, these examples all involve diffusion of a liquid into a solid network, the interaction of which releases of free energy of mixing which gets converted to mechanical energy leading to its deformation and swelling of the network. Beyond this simple picture, more complications arise because of several non-linear effects that set in. For example, swelling of the network leads to change in effective deformability of the gel, which in turn affects its porosity and resultant transport of the liquid into the network (Tanaka *et al.* 1979, Hong *et al.* 2008, Hong *et al.* 2009). In many cases the solid consists of layered structures with different stiffness and swellability, so that geometric and material heterogeneity results in localised bending and curling (Dawson *et al.* 1997, Kim *et al.* 2010). The combined effect of geometry, solid-liquid interaction and their physical properties allow such a sheet to attain shapes, which are not achieved in conventional methods (Lee *et al.* 2012). While, these are all examples of static equilibrium which causes deformation but no displacement or locomotion of solid, recently we have shown that for an axisymmetric elastomeric cylinder, simultaneous diffusion of a solvent into it and its evaporation, results in a dynamic equilibrium: the cylinder does not simply bend and straighten out, but also rolls on a substrate (Hore *et al.* 2012). The rolling motion continues as long the cycle of solvent diffusion and evaporation continues to occur. The driving force for rolling can be strong enough that a cylinder can drag cargo of 8-10 times its own body weight and can also roll up an inclined plane.

The objective of this report is to show that rolling motion is linked to the curvature of the cylinder. The effect is very much like marangoni flow in liquid. In marangoni effect, curvature of a liquid surface evolves following gradient in surface tension, caused either by adsorption of surface active molecules or by temperature. Liquid flow occurs as the curvature "flows" (Krechetnikov 2010) towards a direction in which the surface energy minimizes. Similar to that, for a solid cylinder, asymmetric swelling results in finite curvature of bending which first increases eventually exceeding a threshold value at which the static state of the cylinder bifurcates to the dynamic state of rolling. Beyond it, the curvature attains a plateau value and remains so, as long as the cylinder continues to be powered by the solvent; during this phase, the rolling velocity of the cylinder remains unaltered. We show that the balance of bending energy with the kinetic energy and several modes of energy dissipation yields an expression of the rolling velocity as a function of several material and geometric properties of the solid and the liquid. This scaling relation captures most of the observations in experiments including that rolling velocity increasing with temperature of the substrate and the surrounding, similar to marangoni flow in liquid.

**Experimental:**

**Materials:** Poly(dimethylsiloxane) (sylgard 184, Dow corning product) (PDMS) were used for the preparation of elastomeric micro-cylinder; organic liquids: chloroform, hexane, toluene, heptane, TEA, perfluorooctane, methanol, and acetone, procured from

S.D. fine-chem limited were used as the solvent. Cylindrical shells of steel of different diameter ($450-1200\,\mu m$) procured from local market were used as templates for preparing the samples. Microscope glass slides were coated with monomolecular layers (SAM) of octadecyltrychlorosilane (OTS), used as substrates for carrying out the experiment.

**Method of preparation of sample:** Cylindrical rods of PDMS were prepared by curing a mixture of PDMS precursor liquid and a crosslinker at 10:1 w/w inside the steel shells at 80$^{o}$C for 2 hours. After curing, the PDMS cylinders were pulled gently out and were cut into pieces of length, $l=10\,mm$ using surgical blade.

**Rolling experiment at controlled temperature of the substrate and surroundings:** Figure 1(a) shows the schematic of the experiment in which a cylinder was gently placed on an OTS coated microscope glass slide. The rough surface of the cylinder allowed it to adhere only weakly to the slide. The glass slide was placed on a laboratory hot plate, so that its temperature could be varied from $25-80\,^{o}C$. The temperature of the surroundings at the immediate vicinity of the slide was found to be similar as that of the substrate. Small quantity of a solvent, 0.5- 2µl was dispensed at one side of the cylinder using a micropipette. Solvents having different boiling points, swellability of the network, and surface tensions, were used for the experiment (Lee *et al.* 2003). The rolling motion of the cylinder was video recorded at 60 – 5000 fps for estimating the evolution of curvature of the cylinder and the rolling velocity.

**Results and Discussion:**

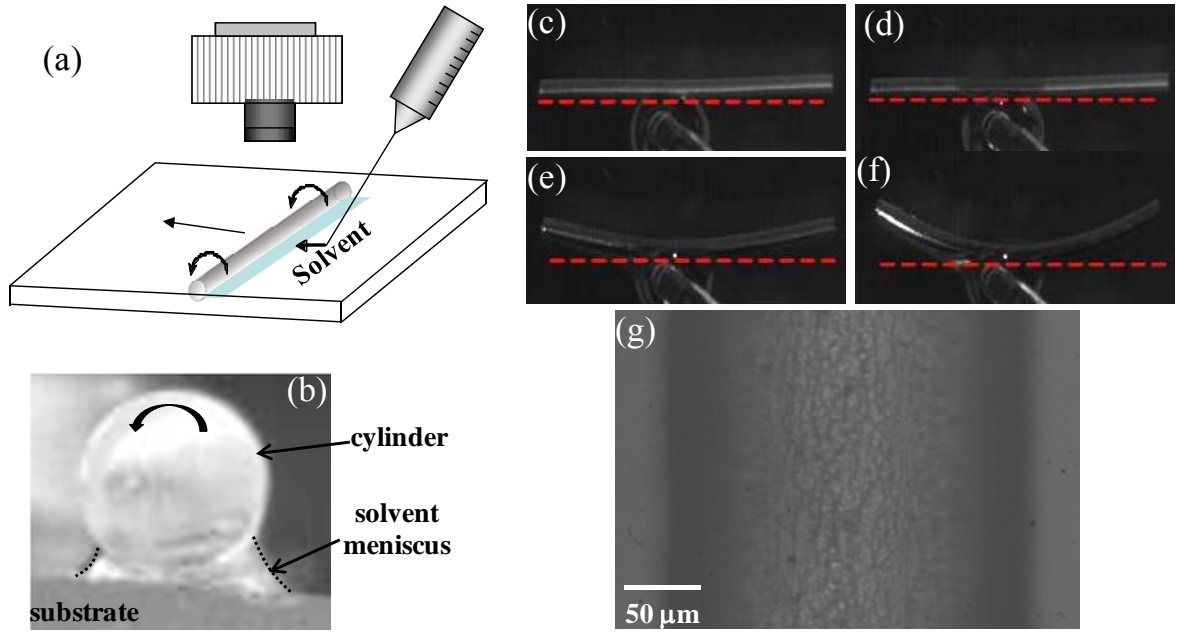

**Figure 1: (a) Schematic of the experimental set up. Here a PDMS cylinder is kept on a microscope glass slide and in its vicinity, desired volume of a solvent is released using a micropipette. (b) Side view of the cylinder depicts accumulation of solvent at its rear side. (c-f) The sequence of video-micrographs, captured using a high frame camera at 6000 fps, depict the change in morphology of the cylinder before it begins to roll. A PDMS cylinder of diameter $d = 340$ μm and length $l = 10$ mm is used in this experiment; $0.5$ μl of heptane is used as solvent. Optical micrograph (c) represents the initial state of the undeformed cylinder before release of the solvent and micrographs (d-f) represent time $t = 0$, $0.045$ and $0.055$ seconds respectively. (g) Optical micrograph of the area of contact of the cylinder with a glass substrate, before the solvent is released.**

**Bending curvature of cylinder:** The sequence of optical micrographs in Fig. 1(c-f) depicts the evolution of a typical PDMS micro-cylinder prior to its rolling. Here a cylinder of diameter, $d = 340$ μm and length, $l = 10$ mm was placed on a substrate kept inclined at an angle 11° with the horizon and maintained at 27° C. 0.5 µl of heptane was released which instantaneously wets the surface of the PDMS and spreads uniformly along the length of it (Fig. 1(d,e)) within 10 m-sec. However, this quantity of solvent is insufficient to completely submerge the cylinder, as a result, it forms a meniscus and wets only a portion of the cylinder. Figure 1(b) shows the front view of the cylinder cross-section, whereas figure 1(g) shows the portion of the contact zone between the cylinder and the substrate. The rough surface of the cylinder at zero contact load condition results in a negligibly small contact area, the contact zone remains discontinuous and porous, so that the solvent can occupy the porous space and form two menisci at the two sides of it (figure 1(b)). Followed by wetting, the solvent diffuses into the network, extent of diffusion over time depends on the diffusivity of the solvent into it. The diffusion coefficient of the solvents can be estimated by using the equation (Tanaka $et$ $al.$ 1979, Holmes $et$ $al.$ 2011) $D = EK/\eta$ where $E$ represents the elastic modulus of the network, $K$ represents its permeability and $\eta$ is the viscosity of solvent respectively. Putting representative numbers, $E = 3 \times 10^6$ Pa, $K = 10^{-18}$ m², and $\eta \sim 0.294$ mPa sec, the diffusion coefficient can be calculated as $D = 10^{-8}$ m²/sec. Using the diffusion coefficient, an estimate of the penetration depth into the network can be obtained as, $h = \sqrt{Dt}$.

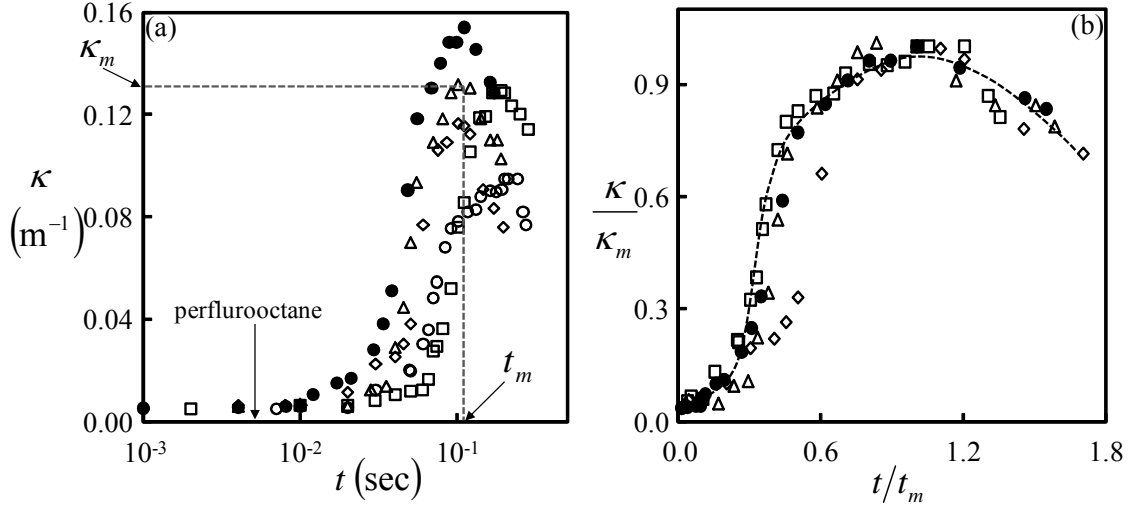

**Figure 2:** (a) The curvature $\kappa$ of cylinder of diameter $d = 340$ μm is plotted against time $t$ following initial release of 0.5 μl of solvent in its vicinity. Symbols ◇, △, ●, ○, and □ represent solvents hexane, heptane, TEA, chloroform and toluene respectively. (b) The curvature and time data are scaled respectively with maximum curvature $\kappa_m$ and the time $t_m$ within which the cylinder attains the maximum curvature. The scaled curvature $\kappa/\kappa_m$ plotted against $t/t_m$ for different solvents, all collapse on a single master curve.

Contact of solvent with the elastomeric network releases excess free energy of mixing and extent of energy release depends on their compatibility expressed in terms of solubility parameter of each material as (Flory 1942): $\delta = (\Delta_{vap} U/V)^{1/2}$, in which the term within parenthesis defines the cohesive energy density. More accurate representation of solubility parameter demands polar and non-polar interactions to be separately considered. Nevertheless, species having similar values of $\delta$ are more compatible and are therefore easier to solubulize than others with widely separated $\delta$ values. Table 1, shows that for PDMS, $\delta_{PDMS} = 7.3$, while that for solvents: hexane, heptane and triethylamine, $\delta$ values are 7.3, 7.4 and 7.5 respectively, suggesting that hexane is the most compatible solvent for PDMS. In the context of mixing of a solvent and a polymeric network, the free energy change due to their interaction is written as (Flory 1942), $\Delta_{mix}\underline{G} = \underline{V_m}(\delta_1 - \delta_2)^2 \phi_1\phi_2 + RT\sum_{i=1}^{2} x_i \ln x_i$, in which $x_i|_{i=1,2}$ are the mole fractions of solvent and the solid network respectively and the first term in right hand side defines the enthalpy change of mixing, $\Delta_{mix}\underline{H}$. The above relation suggests that $\Delta_{mix}\underline{H}$ diminishes for species which are compatible, finally vanishing for ones like PDMS and hexane. Vanishing enthalpy also implies an ideal solution which results in maximization of the free energy of mixing per mole of mixture: $\Delta_{mix}\underline{G} = RT(x_{solvent} \ln x_{solvent} + x_{solid} \ln x_{solid})$. Since, $x_{solid}(r,t) = 1 - x_{solvent}(r,t)$, mixing energy released per unit volume can be written as, $\Delta_{mix}G_V = NRT(x_{solvent} \ln x_{solvent} + (1 - x_{solvent})\ln(1 - x_{solvent}))$, in which $N$ represents the moles per unit volume of the cylinder.

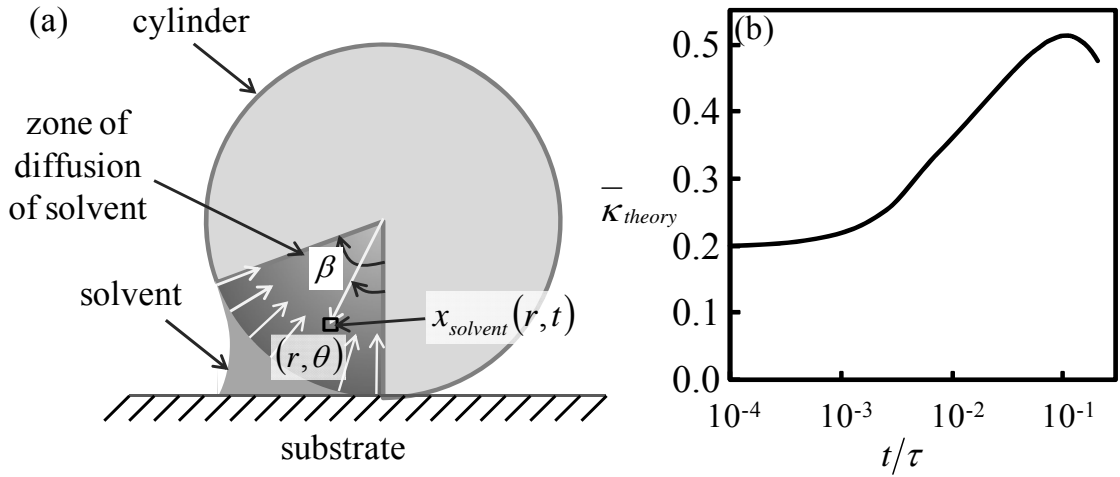

**Figure 3. (a) Schematic of cylinder cross-section in contact with a pool of solvent on a substrate. The solvent is assumed to diffuse radially into the cylinder within the shaded region. The mole-fraction of solvent varies with radial distance from the central axis of the cylinder remaining uniform along its length. (b) Normalized curvature, $\overline{\kappa}_{theory}$ as calculated by using equation 2 is plotted against dimensionless time $t/\tau$.**

An estimate of the mole fraction of solvent $x_{solvent}(r,t)$ at any location within the cylinder can be obtained by solving the diffusion equation under the assumption that (a) solvent diffuses in radial direction within the shaded area as in figure 3(a) and (b) solvent concentration remains uniform along the length of the cylinder. The detailed solution of the simplified equation along with the boundary conditions that at $r = d/2$, $x_{solvent} = 1$ and at $r = 0$, $dx_{solvent}/dr = 0$ has been presented in Appendix I, which yields the following relation for the mole fraction of solvent,

$$x_{solvent}(r,t) = 1 - 2\sum_{m=1}^{\infty} \frac{J_0\left(\frac{2\alpha_m r}{d}\right)}{\alpha_m J_1(\alpha_m)} e^{-\left(\frac{2\alpha_m}{d}\right)^2 Dt} \quad \text{at } 0 < \theta < \beta \text{ and } 0 < z < l \quad (1)$$

Here $J_0(s)$ and $J_1(s)$ are the Bessel functions of order zero and one respectively. The constant, $\alpha_m$ is the solution of $J_0(\alpha_m) = 0$ and are obtained as, $\alpha_1 = 2.405$, $\alpha_2 = 5.52$, $\alpha_3 = 8.654$, ... respectively. The energy released over the shaded area of the cylinder is then obtained from the following integral,

$$\Delta_{mix} G = NRT \int_0^{d/2} (x_{solvent} \ln x_{solvent} + (1 - x_{solvent}) \ln(1 - x_{solvent}))(l\beta r) dr$$

in which, we can substitute the expression of $x_{solvent}(r,t)$ from equation 1. Assuming that this energy is completely used up in bending the cylinder and noting that bending energy per unit length of the cylinder is given as, $EI\kappa^2/2$, where $EI$ is the bending rigidity of the cylinder, the relation for curvature is deduced as,

$$\kappa = \sqrt{\frac{2N\beta RT}{EI}} \left[ \int_0^{d/2} (x_{solvent} \ln x_{solvent} + (1 - x_{solvent}) \ln(1 - x_{solvent})) r dr \right]^{1/2} \quad (2)$$

The semi-log plot in figure 3(b) shows the normalized curvature, $\bar{\kappa}_{theory}$ as a function of dimensionless time $t/\tau$ in which diffusion time $\tau$ is defined as, $\tau = d^2/4D$. The data suggest that the swelling induced curvature of the cylinder is expected to vary non-monotonically with times as was obtained also for more symmetric cases like swelling of a thin rectangular strip. For $d = 340$ μm and $D = 10^{-8}$ m$^2$/sec, $\tau$ can be calculated as, 2.89 sec, so that the time for attainment of maximum curvature, according to figure 3(b), is ~0.3 sec. Somewhat more accurate an estimation of this time can be made by solving the diffusion equation along both $r$ and $\theta$ co-ordinates. Nevertheless, the result of non-motonic variation of curvature with time was examined by using liquids of different types, e.g. liquids which diffuse into the PDMS network to a very small or insignificant extent (Lee *et al.* 2003), e.g. perfluorooctane, acetone ($\delta = 9.9$), or methanol ($\delta = 14.5$) and also with the ones, like hexane, toluene, heptanes and chloroform which are known solvents for the elastomer with similar swelling ratios (Lee *et al.* 2003). The former ones were observed neither to swell the cylinder nor to deform it or drive it to roll. For the later ones, the cylinder was observed to deform, the extent of deformation depending on several parameters like length and diameter of cylinder, swelling ratio of the solvent and so on. In Figure 2(a), we plot the evolution of curvature of a typical cylinder of diameter $d = 0.34$ mm, with time for several solvents; here, $t = 0$ defines the time at which the solvent completely spreads along the length of the cylinder following which it begins to diffuse into the network. For each case, the curvature was found to increase from zero to a maximum value, $\kappa_m$ at time, $t_m$, beyond which it decreased to finally attain a plateau value, thus corroborating with the plot in figure 3(b). $\kappa_m$ and $t_m$ were different for different solvents e.g. for hexane with solubility parameter, $\delta = 7.3$ and diffusivity, $D = 10^{-8}$ m$^2$/sec, a cylinder of diameter $d = 0.34$ mm attained $\kappa_m = 0.12$ mm$^{-1}$ within $t_m = 0.083$ sec. However, for toluene, with $\delta = 8.9$ and $D = 0.54 \times 10^{-8}$ m$^2$/sec, curvature $\kappa_m = 0.11$ mm$^{-1}$ was attained over $t_m = 0.12$ sec. It is to be noted that for either solvent, $t_m$ was found to be smaller than that predicted in figure 3(b) because initiation of rolling motion limits the maximum curvature that is reached. The time $t_m$ for attainment of maximum curvature $\kappa_m$ of the cylinder defines a timescale for each solvent. It is then expected from relation (1 and 2) that curvature $\kappa$ scaled with maximum curvature $\kappa_m$ for respective solvent be independent of all other parameters except time $t$, scaled with $t_m$. It is indeed observed in figure 2(b) in which $\kappa/\kappa_m$ data for all different solvents plotted against $t/t_m$ collapse onto a single master curve, corroborating with earlier observation (Holmes *et al.* 2011) with bending of swollen PDMS strip. Furthermore, if we consider the whole of $t_m$ as the time through which a solvent diffuses into the network, we get an estimate of the diffusion distance $h_m$ of the solvent. In Table 1, for different solvents, although $t_m$ varies depending on the respective diffusivity, this distance $h_m$ is calculated to be almost same ~ 26 μm, much smaller than the characteristic lengthscale in the experiment, e.g. diameter $d = 340$ μm of the cylinder. In essence, the solvent diffuses only to a very small extent, in contrast to earlier observations on rectangular strips for which solvent diffuses whole through the thickness of the strip; the time of diffusion too remains orders of magnitude larger [Holmes *et al*. 2011].

**Initiation of rolling:** The extent of diffusion of the solvent in our problem is limited by initiation of rolling of the cylinder which is directly linked to the curvature of the cylinder. Asymmetric swelling leads to bending of the cylinder in the plane of the substrate while solvent evaporation from its surface tends to straighten it. Because of bending, the centre of mass of the cylinder moves forward from point A to A' as shown in the schematic presented in figure 4(a). The distance $q_x$ between A and A' is a function of curvature, $\kappa$ of the cylinder and its length $l$ and can be deduced as (Appendix II),

$$\frac{q_x}{l} = \frac{1}{\kappa d}\left(1 - \frac{2}{l\kappa}\sin\left(\frac{l\kappa}{2}\right)\right) \qquad (3)$$

In figure 4(e) we show the distance $q_x/l$ plotted for different diameter and length of cylinders, driven to roll by 2μl of hexane. For large diameter of cylinders, $d > 1.2$ mm, $q_x$ remains insignificantly small so that, the center of mass is immediately restored to its original location following evaporation of solvent. But when $q_x/l$ is large, the cylinder can restore to its original state only by catching up with its center of mass. But as it tends to slide forward on the substrate, it induces frictional resistance at the interface with the substrate, which acts in the opposite direction, resulting in a torque which sets the cylinder to roll. It is necessary, however, that the cylinder bends in the plane of the substrate. For example, a short enough cylinder, e.g. ~ 5 mm, bends out of plane as shown in figure 4(c), eventually straightening out following evaporation of solvent. Thus the cylinder does not get subjected to friction induced torque and it does not roll. On the other hand, a long and thin enough cylinder, diameter ~340 μm and length ~20 mm, does not swell uniformly along length; its curvature varies from convex to concave as shown in figure 4(b) suggesting superposition of several bending modes. Here too, the forward torque remains insufficient to drive the cylinder to roll. The cylinder does not roll also if there is no friction at the interface. This fact was verified by keeping the cylinder on the surface of a pool of water in a petridish and by releasing the solvent as before at one side of the cylinder (figure 4(d)). Here too, the cylinder did initially undergo bending because of non-uniform swelling but no rotation at all, eventually, it swelled uniformly everywhere. In order to quantify the threshold condition for rolling, the dynamic state of the cylinder was monitored as a function of its curvature. In figure 5(a), we plot a typical phase diagram of velocity vs. curvature which shows that curvature increases from zero without any rolling of the cylinder; eventually, a threshold curvature is reached beyond which the cylinder begins to roll. Following this initiation, the curvature slightly decreases and then remains almost constant all-through. Eventually, the cylinder straightens out, so that the curvature decreases below a second threshold and the cylinder stops to roll. Corresponding to curvature, the rolling velocity too increases from the static state of the cylinder, eventually attaining a plateau value, beyond which it diminishes again to zero. The plateau value remains nearly but not exactly constant. Therefore, the average velocity at this plateau regime, as represented by the dashed line in figure S2 and 5(b) is considered for further analysis.

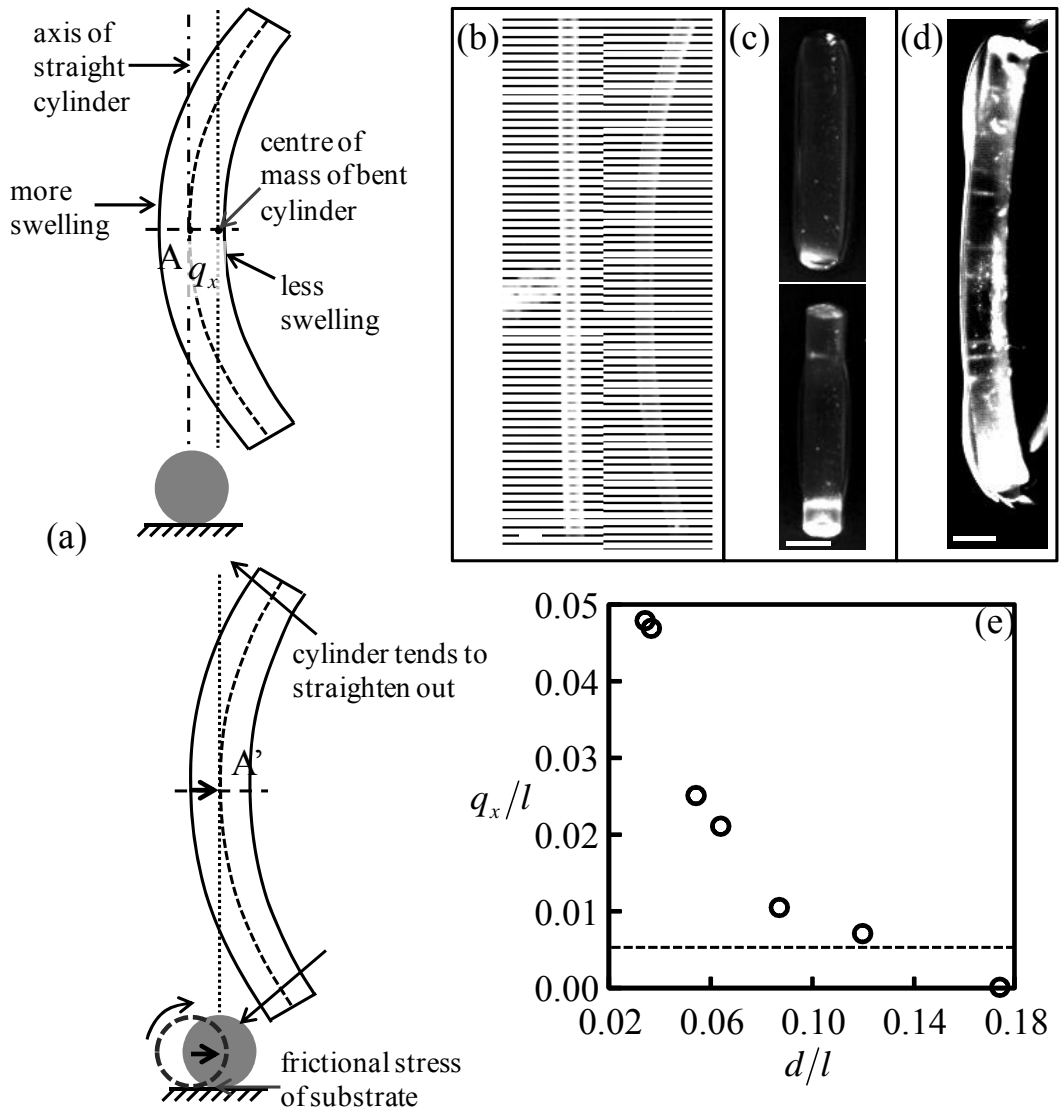

**Figure 4.** (a) Schematic of a cylinder bent due to asymmetric swelling. Its centre of mass moves forward by a distance $q_x$ which depends upon its curvature $\kappa$ and length $l$. The bent cylinder tends to straighten out, in the process, gets subjected to frictional stress on the substrate and begins to roll. (b) Optical micrographs show how a long and straight PDMS cylinder of length, $l = 35$ mm and diameter $d = 870$ μm bends when hexane is released as in figure 1. Because of non-uniform swelling along length, the cylinder does not bend uniformly along length, nor does it roll. (c) Optical micrographs represent a short PDMS cylinder of length $l = 5$ mm and $d = 870$ μm in similar experiment. Here solvent evaporates from two ends

leading to non-uniform swelling along length and cylinder bends out of plane but does not roll. (d) A cylinder of $l = 10$ mm and $d = 870$ μm is gently placed on a pool of water, on which it floats. 2 μl of Hexane is released as in figure 1 which swells the cylinder but does not drive it to roll or swim. (e) The dimensionless distance $q_x/l$ obtained for PDMS cylinders of different diameter and length is plotted against dimensionless diameter $d/l$. In each case 2μl of hexane is used to drive the cylinder. The dashed line shows the threshold limit below which the cylinders do not roll.

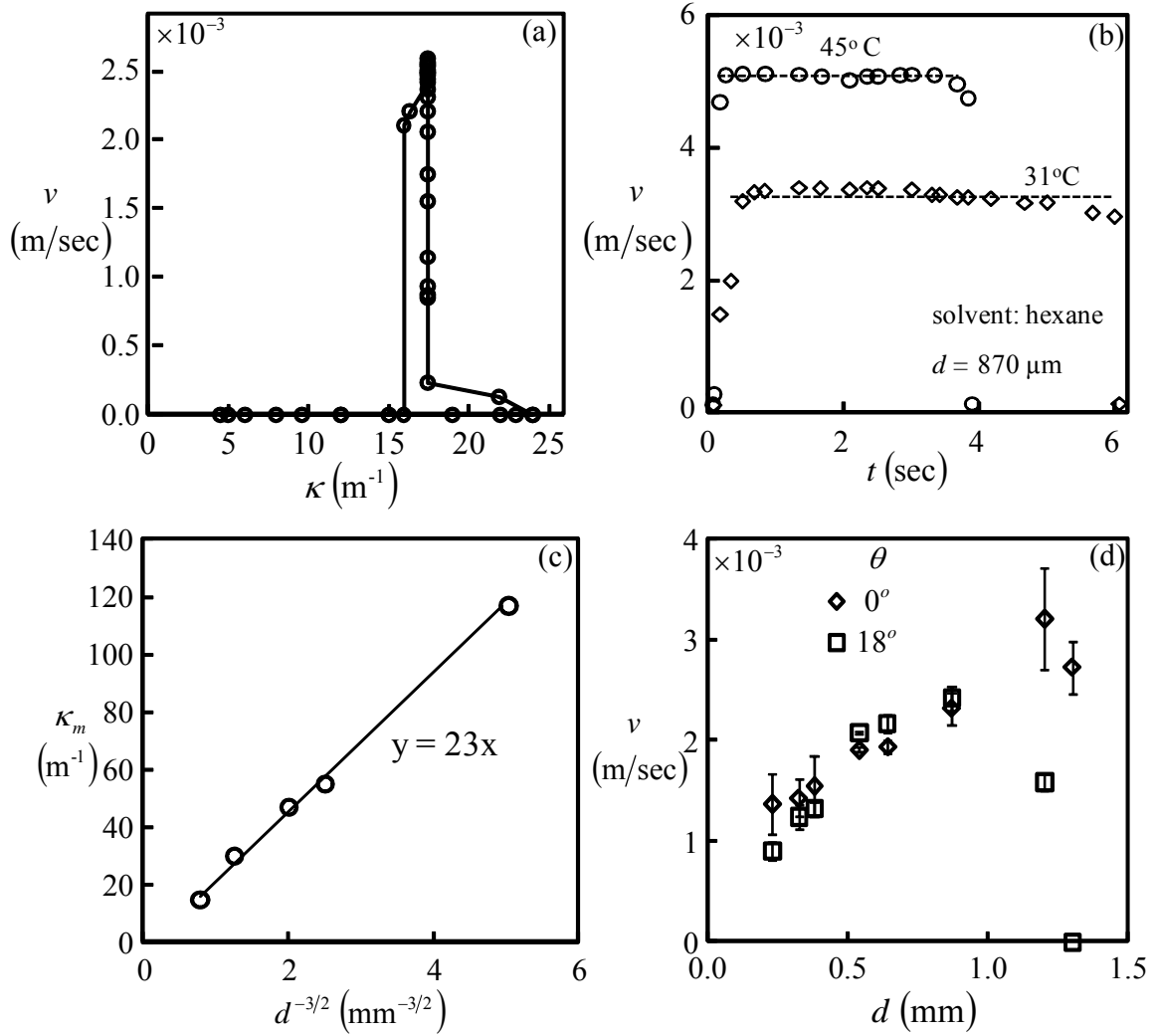

**Figure 5.** (a) A typical bifurcation diagram shows variation of rolling velocity $v$ on a horizontal substrate, of a cylinder of diameter $d = 870$ μm, length $l = 10$ mm and modulus $\mu = 1$ MPa with its curvature $\kappa$ being driven to roll by hexane. (b)

**Variation of rolling velocity $v$ with time $t$ for the experiment depicted in previous plot. The dashed line shows the average velocity considered for further analysis. (c) Experiments with cylinders of different diameter: $d = 0.34 - 1.2$ mm on a horizontal substrate show that the threshold curvature $\kappa_m$ beyond which rolling initiates scales with diameter, as $\kappa_m \sim d^{-3/2}$. (d) Cylinders of different diameter are induced to roll on a horizontal substrate ($\diamond$) and on one kept at an inclination angle $\theta = 18^o$ with the horizon ($\square$). The plot shows that rolling velocity $v$ varies non-monotonically with $d$. These experiments are all carried out by maintaining the substrate temperature at $27^o$ C and using 2 μl of hexane as the solvent to roll.**

The threshold curvature for rolling can be rationalized by balancing two opposite forces that act on the bent cylinder. An estimate of the restoring force per unit length on cylinder due to its bending can be expressed as (Landau *et al.* 1986) $\sim EI\kappa/l^2$, whereas, the frictional resistance is written as $\sim F(mg)$, in which, the quantity within parenthesis is the weight per unit length of the cylinder and $F$ defines the friction coefficient at the interface of cylinder and substrate. Since the contact area between the cylinder and the substrate remains porous and filled with liquid, $F$ depends on properties of the solvent besides that of the solid substrate and the cylinder. In the absence of a detailed knowledge of this dependence, we assume $F$ to be constant and equate the two forces yielding an expression for the threshold curvature for rolling as $\kappa_m \sim (Fl^2 mg/EI) \sim (l/d)^2 (F\rho_s g/E)$. This relation qualitatively corroborates with experimental observations that cylinders with smaller diameter and modulus exhibit a higher threshold curvature beyond which it begins to roll. In order to have more quantitative verification of the scaling relation, experiments were done with different diameter of cylinders $d = 0.33 - 1.3$ mm and length $l = 10$ mm, which were induced to roll by releasing 2 μl of hexane as the solvent. The data in figure

5(c) show that the threshold curvature $\kappa_m$ for different diameters scale as $d^{-3/2}$ thus deviating from the theoretical derivation because of the absence of a suitable relation for friction coefficient $F$. In figure 5(d) we plot the corresponding rolling velocities which suggest that with increase in cylinder diameter, $v$ increases eventually reaching maxima at a threshold value of $d$. Beyond this limit, the rolling velocity was found to decrease, eventually diminishing to zero because of insufficient extent of bending. The results in figure 5 show that the critical curvature for rolling is not an absolute quantity as it depends upon the physical properties of the solvent used for inducing it, the diameter of the cylinder and its modulus. In order to deduce a parameter which absolutely defines the dynamic state of the cylinder it is then imperative to examine the energy balance associated with the rolling of the cylinder, which we discuss next.

**Energy balance of rolling motion in the plateau regime:**

Rolling essentially occurs via release of bending energy of the cylinder as its kinetic energy and several dissipative processes. In the plateau regime, however, the bending curvature and therefore the bending energy does not change, because, over a cycle of rotation, whatever energy is released, is continuously sourced by the mixing energy of solvent-solid interaction. This amount of energy can however be estimated by considering that in one complete cycle the bending curvature would decrease from plateau value $\kappa$ to a hypothetical but considerable smaller value $\kappa_0$ in the absence of solvent to diffuse into the network. The release rate of bending energy per unit length of the cylinder can then be expressed as (Landau *et al.* 1986): $\Pi_{BE} = EI(\kappa^2 - \kappa_0^2)/2\tau$ where $\tau$ is the time for one complete rotation of the cylinder. This time consists of two components: duration over which any specific portion of the cylinder remains in contact with the pool of solvent and that over which it is exposed to the atmosphere for the solvent to get completely evaporated from its surface. Since the former is about an order of magnitude smaller than the later, $\tau$ accounts essentially for evaporation of the solvent. Considering the difference in vapour pressure $p(T)$ of the solvent and its partial pressure $p_0$ in the surroundings, the rate of solvent evaporation per unit length of the cylinder can be expressed as $\dot{m}_s = k(p - p_0)\pi d$. Here $k$ (sec/m) is the mass transfer coefficient which is expected to vary with solvents but for the sake of simplicity is assumed to be independent of it. Although, at any instant only a fraction of the cylinder periphery remains in contact with the solvent and swells it because of solvent diffusion, so that the cylinder swells neither completely nor uniformly, over one complete cycle, the whole of it contacts the solvent. Therefore, considering for any solvent an average swelling ratio $S_r$, the quantity of it that diffuses per unit length into the cylinder over one cycle of rotation can be approximated as, $m_s = \rho_l \pi d^2 (S_r^2 - 1)/4$. Here we have assumed that hypothetical curvature $\kappa_0$ is small enough that the corresponding average swelling ratio nearly equates to 1. Since at steady state the same quantity of solvent dries out of curved surface of the cylinder over a cycle, the characteristic time $\tau$ can be written as $\tau = \rho_l d (S_r^2 - 1)/4k(p(T) - p_0)$.

**Kinetic energy of rolling**: The kinetic energy of a rolling cylinder consists of two components: the translational kinetic energy, $mv^2/2$ and the rotational kinetic energy $I_c \varpi^2/2$, both defined per unit length of the cylinder. Here $I_c$ defines the moment of inertia (Landau *et al.* 1986) about the axis of rotation and $\varpi$ defines the rotational velocity. A straight cylinder rotates about its central axis, therefore $\varpi$ is defined as $\varpi = 2v/d$. For a cylinder bent to a finite curvature, although the centre of mass moves ahead, in the plateau regime, the cylinder exhibits pure rolling without slipping or hopping motion, which suggests that cylinder rolls about its central axis with the rotational velocity approximated as, $\varpi = 2v/d$. Therefore the moment of inertia remains identical to that of the straight cylinder about its central axis and can simply be written as, $I_c = md^2/8$. The kinetic energy expression per unit length of the cylinder is then deduced as $\Pi_{KE} = 3mv^2/4$.

**Energy dissipation:** The dissipative processes consist of two components: adhesion hysteresis at the elastomer-substrate interface and viscous dissipation at the liquid layer sandwiched between these two adherents. The former occurs because the cylinder, while rolling on the substrate, forms a finite contact area which gets continually renewed. At its advancing side fresh area of contact is formed while debonding occurs along the trailing edge. Since, energy associated with debonding, $W_o$ usually exceeds that for crack closure, $W_c$, the hysteretic energy dissipation over one complete cycle can be written as: $\Pi_{hys} = A_{rough}(W_o - W_c)/\tau = A_{rough}W_{hys}/\tau$ in which $A_{rough}$ defines the actual area of contact between the two adherents per unit length of the cylinder. Because of roughness of cylinder surface, $A_{rough}$ remains significantly smaller than what appears for a smooth cylinder. The second component of the dissipative process occurs in this sandwiched layer of liquid which slip on the smooth surface of the substrate but remains adhered to the surface of the PDMS network. As a result, when the cylinder rolls, the liquid at its vicinity moves with a velocity $-v$, whereas that at the vicinity of the solid substrate its velocity is $v$. The velocity gradient of the layer of liquid can then be expressed as: $\dot{\varepsilon} = 2v/\Delta$ where $\Delta = w^2/d$ is the average thickness of the liquid layer. The rate of viscous energy dissipation per unit length of cylinder can then be written as $\sim \eta\left(\dfrac{2v}{\Delta}\right)wv$, where $\eta$ is the viscosity of the solvent. Over a cycle of rotation, the viscous energy dissipation per unit of length of cylinder can be written as $\Pi_{vis} = \eta\left(\frac{2v}{\Delta}\right)wv\tau$. Substituting the expression for $\tau$ and $\Delta$ as derived earlier, $\Pi_{vis}$ can be rewritten as, $\Pi_{vis} = \frac{\eta v^2}{2w}\frac{\rho_l d^2(S_r^2-1)}{k(p(T)-p_0)}$.

**Energy balance:** Since the cylinder remains in thermal equilibrium with the surroundings and the substrate, it is reasonable to assume that the latent heat of vaporization of the solvent is drawn from it. As a result, it does not enter into the following balance of energies which are written for one complete rotation of the cylinder:

$$\frac{EI(\kappa^2 - \kappa_0^2)}{2} = \frac{3mv^2}{4} + AW_{hys} + \frac{\eta v^2 \rho_l d^2(S_r^2-1)}{2wk(p-p_0)} \qquad (4)$$

Putting representative numbers, $d = 0.34$ mm, $E = 3.0$ MPa and $\kappa = 0.12$ mm$^{-1}$, the bending rigidity and bending energy per unit length of the cylinder can be written respectively as $EI = 2.0 \times 10^{-9}$ Nm$^2$ and $\Pi_{BE} = 1.0 \times 10^{-5}$ J. Similarly, putting $v = 1.0$ mm/sec, $l = 10$ mm and $\rho = 1.0$ gm/cc, the kinetic energy per unit length of the cylinder can be written as $\Pi_{KE} = 6.8 \times 1.0^{-11}$ J. The adhesion hysteresis $W_{hys}$ for PDMS used for preparing the cylinder has been found to be $\sim 10 - 20$ mJ/m$^2$ (Chaudhury et al, 1991), while the actual area of contact $A$ is only a fraction $\sim 1\%$ of what would have been achieved for a cylinder with smooth surface. Thus neglecting the $\Pi_{KE}$ and $\Pi_{hys}$, and rearranging equation 4, the rolling velocity is deduced as,

$$v^2 = k\frac{E\pi w(p-p_0)d^2}{64\eta\rho_l(S_r^2-1)}(\kappa^2 - \kappa_0^2) \qquad (5)$$

Experiments with cylinders of different diameter show that width of the sandwiched layer of liquid scales linearly with cylinder diameter, $w \sim d$ (supporting online figure S3);

furthermore, considering that the partial pressure of solvent in the surroundings of the cylinder is negligibly small, $p_0 = 0$ the above equation gets further simplified to,

$$v = \sqrt{k}\sqrt{\frac{E\pi p}{64\eta\rho_l(S_r^2-1)}}(d^{3/2}\kappa)(1-(\kappa_0/\kappa)^2)^{1/2} \qquad (6)$$

in which the last term in parenthesis can be expanded as,

$$(1-(\kappa_0/\kappa)^2)^{1/2} = 1 - \frac{1}{2}(\kappa_0/\kappa)^2 - \frac{1}{8}(\kappa_0/\kappa)^4 + ... \qquad (7)$$

As a first approximation, considering only the first term in equation 7, i.e. considering that bending energy would get completely dissipated in the absence of a source, $\kappa_0 \to 0$ and writing $\xi = \sqrt{\frac{E\pi p}{64\eta\rho_l(S_r^2-1)}}(d^{3/2}\kappa)$, equation 6 can be simplified as,

$$v = \sqrt{k}\xi \qquad (8)$$

Thus the rolling velocity is deduced as a function of $\xi$ which accounts for the effect of all relevant parameters including curvature of the cylinder, its diameter and elastic modulus and all relevant solvent properties.

Dependence of rolling velocity of the cylinder on its curvature is reminiscent of the Marangoni effect of liquid in which convection of liquid occurs from region of higher to lower curvature [Krechetnikov 2010], the gradient in curvature being generated by adsorption of surfactant molecule to the liquid surface or via local temperature gradient. Balancing the surface stress with the viscous stress over a characteristic lengthscale $l$, a scaling relation for liquid velocity can be obtained as, $v \sim (\gamma l/\mu)\kappa$ which is similar to that derived in equation 6 for rolling of solid cylinder. Furthermore, dividing the expression in equation 6 with the shear wave speed, $\sqrt{E/\rho_s}$ of the solid, a dimensionless number can be deduced as $\sqrt{kd}\sqrt{\dfrac{\rho_s}{\rho_l}\dfrac{\pi p}{64\eta(S_r^2-1)}}(d\kappa)$ which is an equivalent of the Marangoni number in liquid. The difference between the two is that the solid begins to roll only beyond a threshold curvature or bending stress whereas for Marangoni effect in Newtonian liquid, infinitesimal difference in surface stress causes liquid to flow.

The validity of equation 8 can be examined in experiments in which curvature of cylinder can be varied by using variety of cylinders and solvents. Use of different solvents and/or cylinders however changes the values of more than one parameter simultaneously, which does not allow their effect to be examined individually. In other word, these parameters do not allow the curvature to be varied over a range, for one particular set of cylinder and solvent. In order to accomplish it two different kinds of experiments were mooted: in one case the cylinder was placed on an inclined plane, the inclination of which was systematically varied; in the second kind of experiments, the temperature of the substrate and the surrounding was varied over a range, from room temperature to over boiling point of that of the solvent. These two types of experiments were carried out for several solvent-cylinder combinations: chloroform, toluene, heptanes and hexane as solvents and cylinders of diameters $d = 0.34 - 0.87$ mm to yield large variety of data.

Figure 6 summarizes the result from a typical set of experiments in which hexane is used a solvent to induce a cylinder of diameter $d = 0.87$ mm and Young's modulus $E = 3.0$ Mpa. As observed previously, for each angle of inclination, a plateau value of curvature is attained, which remains constant almost whole through the rolling of the cylinder. The plateau value of curvature however does not vary monotonically with $\theta$, but attains a maxima at an intermediate angle of $\theta = 11^o$ (inset of figure 6b), at which the rolling velocity too is found to be maximum. When the cylinder rolls up an inclined plane, the solvent accumulates more on its rear side, thereby causing enhanced asymmetry in swelling and consequent larger curvature of bending. Following equation 10, the rolling velocity is then indeed expected to increase with $\theta$. At large $\theta$, however, e.g. $\theta = 35^o$, the effect of gravity becomes strong enough that both the solvent and the cylinder are pulled down the plane, so that cylinder curvature diminishes almost to zero, consequently the cylinder fails to roll. Similar observations were made in all different solvents and as noted previously, these experiments allow the cylinder curvature to be varied without altering any other parameter except the angle of inclination.

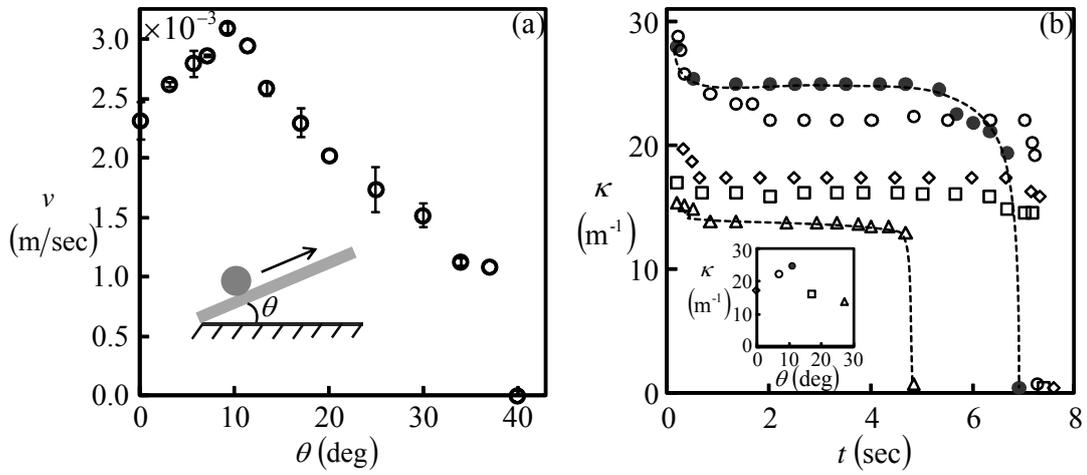

**Figure 6. (a) A cylinder of diameter $d = 0.87$ mm is induced to roll up an inclined plane using hexane as solvent. The rolling velocity $v$ is plotted against the angle of inclination $\theta$ which show that $v$ maximizes at an intermediate $\theta$. (b) The plot shows the evolution of curvature of the cylinder over time for few different $\theta$. Symbols $\Diamond$, $\bigcirc$, $\bullet$, $\square$ and $\triangle$ represent $\theta = 0^o, 7^o, 11^o, 17^o$ and $27^o$ respectively. The inset shows the plateau value of curvature plotted against inclination angle $\theta$.**

The data (open symbols) from different experiments were scaled as in equation 8 and plotted in figure 7, which shows that all data can be fitted to a master curve. It is to be noted that although $S_r$ defines the average swelling ratio as discussed earlier, in the absence of measured data from experiments, equilibrium swelling ratio values as obtained from Lee et al, 2003 were used in generating figure 7. Since for different solvents, the equilibrium swelling ratios are very similar, varying from 1.31-1.39, it is reasonable to think that the error inserted because of using the equilibrium values is similar for all solvents. The slope of the master curve then yields a value of mass transfer coefficient which remains nearly constant irrespective of solvents. Furthermore, finite intercept in the x-axis shows that the second term in right hand side of equation 7 can be small but not negligible, implying that bending energy does not get completely dissipated in one cycle and the parameter $\xi$ has to exceed a threshold $\xi_c \sim 28$ in order for rolling to occur.

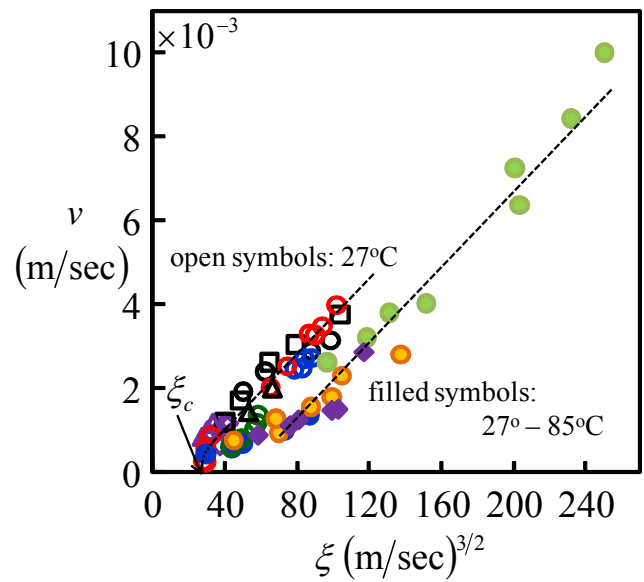

**Figure 7. Rolling velocity of cylinder from large number of experiments is plotted against the quantity $\xi$. Symbols $\bigcirc$, $\square$, $\diamondsuit$ and $\triangle$ represent the data obtained using solvents hexane, heptane, chloroform and toluene respectively. Unfilled symbols represent experiments carried out on substrates kept at room temperature but inclined at different inclination angle: $\theta = 0^o$ to $27^o$ with the horizon. Filled symbols represent second set of experiments done on horizontal substrates but maintained at different temperature: $T = 27^o$ C to $80^o$ C. Cylinders of diameter $d = 0.34 - 0.87$ mm are used in experiments.**

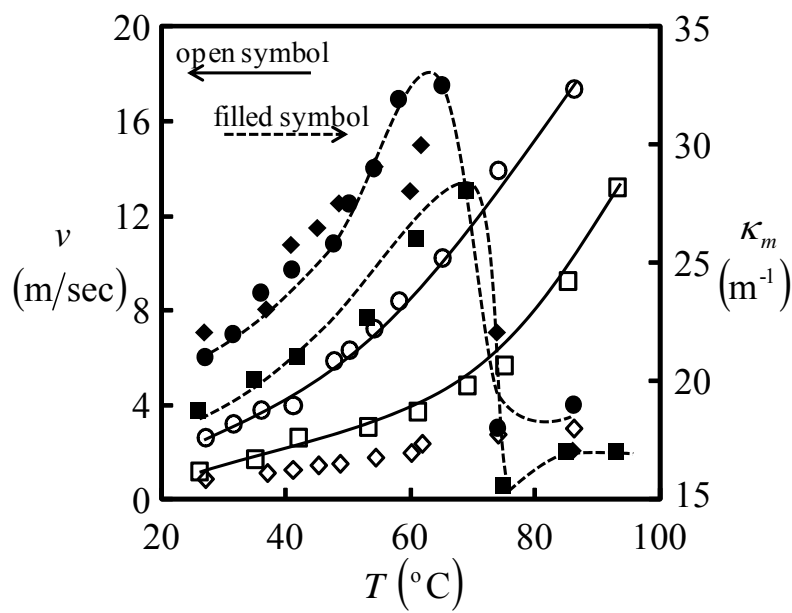

**Figure 8.** A cylinder of diameter $d = 0.87$ mm is made to roll on a horizontal plane heated to different temperature using different solvents. The rolling velocity $v$ and the corresponding plateau value of curvature $\kappa$ are plotted against the substrate temperature $T$. Symbols ○, □ and ◇ represent solvents hexane (b.p. 69° C), heptane (b.p. 98° C) and chloroform (b.p. 60° C) respectively.

**Rolling of cylinder at elevated temperature of substrate and surrounding:** The observation, that a cylinder rolls at higher velocity when driven by a solvent of lower boiling point, motivates us to examine the effect of temperature on rolling. We preheated the substrate and the surroundings to an elevated temperature, at which the rate of diffusion of the solvent into the network enhances. Increase in diffusion of solvent results in increased curvature of the cylinder. In Figure 8 (also in Movie_1 and Movie_2) a cylinder of diameter $d = 870$ μm was placed on glass substrates heated to different temperature: 25 - 75 °C and was then driven to roll by releasing 2 μl of three different solvents: hexane, heptane and chloroform. The maximum curvature $\kappa_m$ of the cylinder was found to vary non-monotonically with temperature suggesting two opposing effects which determine the curvature of the cylinder: increase in diffusion of solvent into the network at higher temperature tends to enhance the asymmetric swelling of the cylinder and consequent bending whereas faster evaporation of it from cylinder surface tends to diminish this asymmetry. Initially, with increase in temperature, the balance of these two effects first gets shifted towards higher curvature, but beyond a threshold temperature, the second effect becomes dominant and the plateau value of curvature diminishes to a smaller value. This non-monotonic variation of curvature with temperature, in itself, calls for a detailed analysis which we keep for future. For now, we attempt to employ equation 10 to analyze our data after accounting for the effect of temperature on vapour pressure, density and viscosity of the solvent. The vapour pressure for each solvent is estimated as a function of temperature using Antoine equation, similarly, the viscosity and density data too are obtained for different temperature using relevant literature value. The filled symbols in figure 8 represent rolling velocity plotted against $\xi = \sqrt{\dfrac{E\pi p(T)}{64\eta(T)\rho_l(T)(S_r^2 - 1)}}\left(d^{3/2}\kappa\right)$, which show that, similar to open symbols, these data too fall on a single master line, albeit slightly away from the open symbols. This discrepancy is possibly due to the dependence of mass transfer coefficient on temperature.

**Summary:**

In summary we can write the following points:

1. In contrast to several known mechanisms of limbless locomotion (Arora *et al.* 2009, Chakrabarti *et al.* 2014, Maeda *et al.* 2008, Mahadevan *et al.* 2004, Yeghiazarian *et al.* 2005) we have presented here a novel mode i.e. rolling of a soft cylinder, powered by small quantity of a solvent. We have shown that bending induced by asymmetric swelling is what drives this motion so that rolling velocity increases when bending curvature of the cylinder is increased. The maximum velocity that is achieved compares with other mechanisms described in literature and can be tuned by varying large number of parameters: diameter of the cylinder, its elastic modulus, density and viscosity of solvent, substrate and surrounding temperature and others.

2. Unlike several other known systems, the mechanism here does not require any moving component or device to drive the cylinder, rather free energy of mixing of the solvent and the polymeric network gets converted to mechanical energy.

3. We have shown that curvature of the cylinder, i.e. its bending energy can be increased by altering various parameters, one of them being the temperature of the surroundings and the substrate. Since, rate of evaporation of the solvent increases with temperature, thereby increasing the asymmetric effect of bending, these observations suggest that the latent heat of vaporization of the solvent is at least partially, if not completely, drawn from the surroundings instead of being provided by the chemical energy of mixing. Although, the exact partitioning of the energy with respect to solvent evaporation is not yet clear, it can be safely concluded that drawing energy from the surrounding enhances the overall efficiency of the rolling mechanism.

4. Some aspects of the rolling motion are rather akin to Marangoni flow in liquid. For example, the role of curvature is similar in both these cases, i.e. to drive the material to flow or to locomote; the driving force can be strong enough that the material can be driven up an inclined plane (Chaudhury *et al.* 1992). In both the cases, the curvature and therefore the velocity remains a function of temperature.

5. Our preliminary experiments suggest that the cylinders can both push and pull a cargo weighing more than its own weight, and unlike several other mechanisms, the cylinder can locomote on a hot surface over large temperature range: $27^{\circ}$-$80^{\circ}$C.

6. All these advantages and examples make this mechanism of limbless locomotion an ideal candidate for further exploration in the general area of soft robotics. In nature, it seems, larvae of insect *Pleurotya ruralis* have already adopted the rolling mode of locomotion in an emergency situation (*Brackburry* 1997). The caterpillar's body is structured such that it allows them to walk slowly on a substrate but when attacked by a predator, it chooses to coil into a wheel which can then escape by rolling at a speed as high as 40 cm/sec.

**Acknowledgement:** SM and AG acknowledge the pioneering studies (Hore et al 2012) on rolling of cylinder that A. Majumder, a former PhD student initiated at IIT Kanpur. AG acknowledges also the Department of Science and Technology, Government of India for financial assistance via grant no. SB/S3/CE/036/2013.

Appendix I

Noting that mole fraction of solvent remains $x_{solvent}(r,\theta,t)$ uniform along length, its variation with time can be estimated by solving Fick's second law of diffusion equation in cylindrical co-ordinate,

$$\frac{\partial x_{solvent}}{\partial t} = D\frac{1}{r}\frac{\partial}{\partial r}\left(r\frac{\partial x_{solvent}}{\partial r}\right) + D\frac{1}{r^2}\frac{\partial^2 x_{solvent}}{\partial \theta^2} \qquad (1)$$

in which, we have assumed that diffusivity, $D$ remains unaltered with time and isotropic with respect to $r$ and $\theta$ co-ordinates. The above equation can be further simplified by considering a situation as in figure AI.1 in which the solvent is assumed to diffuse into the cylinder only through the portion of its outer surface that remains in contact with the solvent; the solvent diffuses radially but uniformly with respect to $\theta$, i.e. diffusion occurs only through the shaded portion as in figure 3(a). While the first part of the assumption remains valid before initiation of rolling motion of the cylinder, the second part is not expected to hold good over long time as the solvent once diffused into the shaded area is expected to diffuse along $\theta$ co-ordinates in regions at the vicinity of it. Over short time, however, this assumption is expected to remain nearly constant so that equation 1 is simplified as,

$$\frac{\partial x_{solvent}}{\partial t} = D\frac{1}{r}\frac{\partial}{\partial r}\left(r\frac{\partial x_{solvent}}{\partial r}\right) \qquad (2)$$

Equation (2) is required to be solved using the following boundary conditions:

At $r=0$, $\left.\dfrac{dx_{solvent}}{dr}\right|_{r=0} = 0$, and $r=d/2$, $x_{solvent}(r=d/2,t)=1$

and initiation condition that $x_{solvent}(r<d/2,0)=0$. We solve equation 2 using the separation of variable method, in which, $x_{solvent}(r,t)$ can be written in terms of eigen functions as, $x_{solvent}(r,t) = 1 - \Phi(r)\Theta(t)$, so that, at $r=d/2$, $\Phi(r)\Theta(t)=0$ and at $r=0$,

$\left. \dfrac{d[\Phi(r)\Theta(t)]}{dr} \right|_{r=0} = 0$. Above substitution for $x_{solvent}(r,t)$ in equation 2 yields the following result, the right hand side of which is a constant:

$$\frac{1}{\Theta(t)}\frac{1}{D}\frac{\partial \Theta(t)}{\partial t} = \frac{1}{r\Phi(r)}\frac{d}{dr}\left(r\frac{d\Phi(r)}{dr}\right) = -\left(\frac{2\alpha_m}{d}\right)^2$$

From above expression, the solutions of $\Theta_m(t)$ and $\Phi_m(r)$ are obtained as,

$\Theta_m(t) = A_m e^{-\left(\frac{2\alpha_m}{d}\right)^2 Dt}$ and $\Phi_m(r) = J_0\left(\dfrac{2\alpha_m r}{d}\right)$ respectively, in which $J_0\left(\dfrac{2\alpha_m r}{d}\right)$ is the Bessel function of order zero. Substitution of the solutions for $\Theta_m(t)$ and $\Phi(r)$ in expression for $x_{solvent}(r,t)$ and summation over all different eigen modes results in the following form,

$$1 - x_{solvent}(r,t) = \sum_{m=1}^{\infty} A_m J_0\left(\frac{2\alpha_m r}{d}\right) e^{-\left(\frac{2\alpha_m}{d}\right)^2 Dt} \quad (3)$$

In which the amplitude $A_m$ for eigen modes can be obtained from orthogonality condition,

$$A_m = \frac{8}{d^2 J_1^2(\alpha_m)} \int_0^{d/2} r J_0\left(\frac{2\alpha_m r}{d}\right) dr = \frac{2}{\alpha_m J_1(\alpha_m)} \quad (4)$$

Substitution of the expression for $A_m$ in equation 3 finally results in the evolution of radial distribution of solvent concentration over time:

$$x_{solvent}(r,t) = 1 - 2\sum_{m=1}^{\infty} \frac{J_0\left(\dfrac{2\alpha_m r}{d}\right)}{\alpha_m J_1(\alpha_m)} e^{-\left(\frac{2\alpha_m}{d}\right)^2 Dt} \quad (5)$$

Since, $\Phi_m(r = d/2) = 0$, solution of $J_0(\alpha_m) = 0$ yields different value of $\alpha_m$ as, $\alpha_1 = 2.405$, $\alpha_2 = 5.52$, $\alpha_3 = 8.654$.

Appendix-II

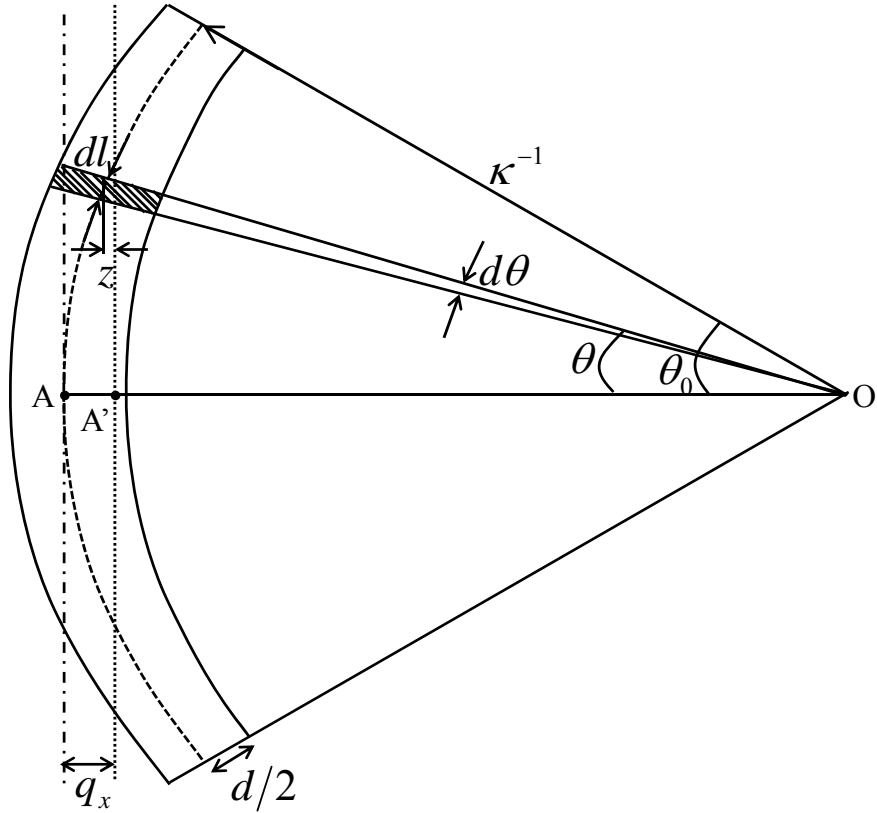

**Figure AII.1. Schematic of a bent cylinder**

Figure $A_1$ shows the schematic of a cylinder of diameter $d$ and length $l$ whose axis and centre of mass goes through point A when it remains straight; however when it is bent to a radius of curvature $\kappa^{-1}$ with centre at O, its centre of mass moves to A'. As a result, the bent cylinder rotates not about an axis that passes through A, but one at A'. The location of A' with respect to A, $q_x$ can be obtained by considering the following integral,

$$\kappa^{-1} - q_x = \frac{4}{\pi d^2 \rho l} \int_{-\theta_0}^{\theta_0} \left(\frac{\pi d^2 \rho}{4} dl\right) \kappa^{-1} \cos\theta = \frac{1}{l} \int_{-\theta_0}^{\theta_0} \left(\kappa^{-1} d\theta\right) \kappa^{-1} \cos\theta \qquad (1)$$

Noting that $\theta_0 = l\kappa/2$,

$$q_x = \kappa^{-1} - \frac{2\kappa^{-2} \sin\theta_0}{l} = \kappa^{-1}\left(1 - \frac{2}{l\kappa}\sin\left(\frac{l\kappa}{2}\right)\right) \qquad (2)$$